\documentclass[sigconf]{acmart}

\usepackage{booktabs} 
\usepackage{enumitem}
\usepackage{xcolor}
\usepackage{xspace}
\usepackage[framemethod=TikZ]{mdframed}

\definecolor{oldlace}{rgb}{0.99, 0.96, 0.9}
\newmdenv [ %
 skipabove=\topsep,
 skipbelow=\topsep,
 leftmargin       = 2              ,
 rightmargin      = 2              ,
 splittopskip     = \topskip      ]{mh}

\newmdenv [ %
 skipabove=\topsep,
 skipbelow=\topsep,
 roundcorner      = 5pt            ,
 leftmargin       = 2              ,
 rightmargin      = 2              ,
 backgroundcolor  = oldlace        ,
 innertopmargin   = 3       ,
 splittopskip     = 3      ]{mq}

\newcommand{\cat}[1]{{\textbf{\emph{#1}}}}
\newcommand{\cc}{\cat{collaborative culture}\xspace}

\begin{document}

\copyrightyear{2018}
\acmYear{2018}
\setcopyright{othergov}
\acmConference[ESEM '18]{ACM / IEEE International Symposium on Empirical Software Engineering and Measurement (ESEM)}{October 11--12, 2018}{Oulu, Finland}
\acmBooktitle{ACM / IEEE International Symposium on Empirical Software Engineering and Measurement (ESEM) (ESEM '18), October 11--12, 2018, Oulu, Finland}
\acmPrice{15.00}
\acmDOI{-}
\acmISBN{978-1-4503-5823-1/18/10}

\title{Building a Collaborative Culture: A Grounded Theory of Well Succeeded DevOps Adoption in
  Practice}

\author{Welder Pinheiro Luz}
\affiliation{%
  \institution{Brazilian Federal Court of Accounts}
  \city{Bras\'{i}lia}
  \country{Brazil}
}
\email{welder.luz@tcu.gov.br}

\author{Gustavo Pinto}
\affiliation{
	\institution{Federal University of Par\'a}
	\city{Bel\'em}
	\state{Par\'a}
}
\email{gpinto@ufpa.br}

\author{Rodrigo Bonif\'acio}
\affiliation{%
  \institution{University of Bras\'{i}lia}
  \city{Bras\'{i}lia}
  \country{Brazil}
}
\email{rbonifacio@cic.unb.br}

\renewcommand{\shortauthors}{W. Luz, G. Pinto, R. Bonif\'acio}
\renewcommand{\shorttitle}{Building a Collaborative Culture: A GT of Well Succeeded DevOps Adoption in Practice}

\begin{abstract}

  \emph{Background.} DevOps is a set of practices and cultural values
  that aims to reduce the
  barriers between development and operations
  teams. Due to its increasing interest and imprecise
  definitions, existing research works have tried to
  characterize DevOps---mainly using a set of concepts and related practices.

  \emph{Aims.} Nevertheless, little is
  known about the \emph{practitioners practitioners' understanding}
  about successful paths for DevOps adoption. The lack of such understanding
  might hinder institutions to adopt DevOps practices. Therefore, our goal
  here is to present a theory about DevOps adoption, highlighting the
  main related concepts that contribute to its adoption in industry.

  \emph{Method.} Our work builds upon Classic Grounded Theory. We interviewed practitioners
  that contributed to DevOps adoption in 15 companies from different
  domains and across 5 countries. We empirically evaluate our model through
  a case study, whose goal is to increase the maturity level of
  DevOps adoption at the Brazilian Federal Court of Accounts,
  a Brazilian Government institution.

  \emph{Results.} This paper
  presents a model to improve both the understanding and guidance
  of DevOps adoption. The model increments the existing view of
  DevOps by explaining the role and motivation of each
  category (and their relationships) in the DevOps adoption process.
  We organize this model in terms of \emph{DevOps enabler categories} and
  \emph{DevOps outcome categories}. We provide evidence that
  \emph{collaboration} is the core DevOps concern, contrasting with an existing
  wisdom that implanting specific tools to \emph{automate building, deployment,
  and infrastructure provisioning and management} is enough to achieve DevOps.

  \emph{Conclusions.} Altogether, our results contribute to (a) generating
  an adequate understanding of DevOps, from the perspective
  of practitioners; and (b) assisting other institutions in the
  migration path towards DevOps adoption.
\end{abstract}

\ccsdesc[500]{Software and its engineering~Collaboration in software development}
\ccsdesc[300]{Software and its engineering~Software creation and management}

\keywords{DevOps, Grounded Theory, Software Development, Software Operations.}

\maketitle

\section{Introduction} \label{sec:introduction}

DevOps is a a set of practices and cultural values that has emerged in the
software development industry. Even before
the existence of the term --- a mix of ``development'' and ``operations''
words~\cite{httermann2012devops} --- companies like Flickr~\cite{flickr}
had already pointed out the need to break the existing separation between
the operations and software development teams. Since then, the term
has appeared without a clear delimitation and gained strength and interest
in companies that perceived the benefits of applying agile practices in
\emph{operation tasks}.
DevOps claimed benefits include increased organizational IT
performance and productivity, cost reduction in software lifecycle, improvement
in operational efficacy and efficiency, better quality of software products and
greater business alignment between development and operations
teams~\cite{characterizing_devops_sbes_2016,state_of_devops,DevOps_Adoption_Benefits_and_Challenges}.
However, the adoption of DevOps is still a challenging task, because there is a
plethora of information, practices, and tools related to DevOps, but it is still unclear
how one could leverage such rich, yet scattered, information in an organized and
structured way to properly adopt DevOps.

Existing research works have proposed a
number of DevOps characterizations, for instance, as a set of concepts with
related
practices~\cite{cooperation_dev_ops_esem_2014,devops_a_definition_xp_15,dimensions_of_devops_xp_15,extending_dimensions_icsea_16,characterizing_devops_sbes_2016,qualitative_devops_journalsw_17}. Although some
of these studies leverage qualitative approaches to gather practitioners' perception (for instance,
conducting interviews with them), they focus on characterizing DevOps,
instead of providing recommendations to assist on DevOps adoption. Consequently,
our {\bf research problem} is that the obtained DevOps characterizations allow a
comprehensive understanding of the elements that constitute DevOps, but do not
provide detailed guidance to support newcomers interested in adopting DevOps.
As a consequence, many practical and timely questions still remain open, for
instance: (1) Is there any recommended path to adopt DevOps? (2) Since
DevOps is composed by multiple elements~\cite{dimensions_of_devops_xp_15}, do
these elements have the same relevance, when adopting DevOps?
(3) What is the role played by elements such as measurement, sharing, and automation
in a DevOps adoption? To answer these questions, we need a holistic
understanding of the paths followed in successful DevOps adoptions.

In this paper, we present a model based on the perceptions of practitioners from
15 companies across five countries that successfully adopted DevOps. The model
was constructed based on a classic Grounded Theory (GT) approach,
and make clear that practitioners interested in adopting DevOps should focus on building a
\cc, which prevents common pitfalls related to focusing on tooling or automation.
We instantiated our model in the Brazilian Federal Court of
Accounts (hereafter TCU), a Brazilian Federal Government institution. TCU was
bogged down in implanting specific DevOps tools, repeating the same non-DevOps
problems, with conflicts between development and operations teams about how to
divide the responsibilities related to different facets in the intersection
between software development and software provisioning. When instantiated,
our model helped TCU to change its focus to improve the collaboration between teams, and to use the tooling
to support (rather than being the goal of) the entire process.
The main contributions of this paper are the following:

\begin{itemize}
\item A model, based on the classic Grounded Theory approach, that could support practitioners interested in adopting DevOps,
      based on evidence acquired from their industry peers;
\item An instantiation of this model in a real world, non-trivial context. TCU is different from the typical \emph{tech companies}
that have successfully reported the adoption of DevOps, though the use of our model there have brought several benefits and
now DevOps practices have been disseminated at TCU.
\end{itemize}

\section{Research Method} \label{sec:research_method}

We used Grounded Theory (GT) as the research method. GT was
originally proposed by Glaser and Strauss~\cite{glase1967discovery}.
As distinguishing features, it has (1) the absence of clear research hypothesis upfront
and (2) limited exposure to the literature at the beginning of the research. GT
is a theory-development approach (the hypothesis emerge as a result of
a investigation), in contrast with more traditional
theory-testing approaches~\cite{coleman2007using}---e.g., those that
use statistical methods to either confirm or refute pre-established hypothesis.

We used GT as the research method due to three main reasons. First, GT is a consolidated
method in other areas of research - notably medical
sociology \cite{gt_medical_sociology}, nursing \cite{barnsteiner2002using}, education
\cite{gt_education} and management \cite{gt_management}. GT is also being increasingly employed
to study software engineering topics~\cite{hoda2017becoming,stol2016grounded,adolph2011using}. Second,
GT is considered an adequate approach to answer research questions that aims to
characterize scenarios under a personal perspective of those
engaged in a discipline or activity~\cite{barnsteiner2002using},
which is exactly the scenario here: what are the successful adoption paths for DevOps? Finally,
GT allows researchers to build an independent and original understanding,
which is adequate to collect empirical evidence directly from the
practice on industry without bias of previous research. The evidence
is only reintegrated back with the existing literature after the step of
theory construction.

Since the publication of the original version of GT~\cite{glase1967discovery},
several modifications and variations have been proposed to the method, coming to
exist at least seven different versions~\cite{denzin2007grounded}.
Here we chose the classic version, mainly because we did not have a research
question at the beginning of our research, exactly as suggested in this
version. We actually started from an area of interest: successful DevOps adoption
in industry. In addition, research works in software engineering that leverage GT
predominantly use this version~\cite{stol2016grounded}.
We carried out our research using an existing
guideline about how to conduct a
Grounded Theory~\cite{adolph2011using} research. This guideline organizes
a GT investigation in 3 steps: \emph{Open Coding} Data Collection,
\emph{Selective Coding} Data Analysis, and \emph{Theoretical Coding}.

\begin{enumerate}[label=(\Alph*)]
\item {\bf Open Coding Data Collection.} We started our research
  by collecting and analyzing data from companies that claim to have
  successfully adopted DevOps.
  To this end, we have conducted a \emph{raw data analysis} that searches for patterns of
  incidents to indicate concepts,  and then grouped these concepts into
  categories~\cite{stol2016grounded}.

\item {\bf Selective Coding Data Analysis.} In the second step, we evolve
  the initial set of
  categories by comparing new incidents with the previous ones. Here the goal is
  to identify a ``core category''~\cite{stol2016grounded}.
  The core category is responsible for enabling the integration of the other
  categories and structuring the results into a dense and consolidated grounded
  theory~\cite{jantunen2014using}. The identification of the core category
  represents the end of the open-coding phase and the beginning of the selective coding.
  In selective coding, we only considered the specific variables that are directly
  related to the core category, in order to enable the production of an harmonic
  theory~\cite{coleman2007using,hoda2011impact}. Selective coding ends when we
  achieve a theoretical saturation, which occurs when the last few
  participants provided more evidence and examples but no new concepts or
  categories~\cite{glase1967discovery}.

\item {\bf Theoretical Coding.} After saturation, we built a theory that
explains the categories and the relationships between the categories.
Additionally, we reintegrated our theory with the existing literature, which allowed us to compare our proposal
 with other theories about DevOps. That is, using a Grounded Theory approach,
 one should only conduct a literature review in later stages of a research,
in order to avoid external influences to conceive a theory~\cite{adolph2012reconciling}.

\end{enumerate}

Throughout the process, we wrote memos capturing thoughts and analytic
processes; the memos support the emerging concepts, categories, and their
relationships~\cite{adolph2012reconciling}.

Regarding data collection, we conducted semi-structured interviews with 15 practitioners of companies from
Brazil, Ireland, Portugal, Spain, and United States that
contributed to DevOps adoption processes in their companies. Participants
were recruited by using two approaches: (1) through direct contact in a \emph{DevOpsDays}
event in Brazil and (2) through general
calls for participation posted on DevOps user groups, social networks,
and local communities. In order to achieve a heterogeneous perspective
and increase the wealth of information in the results,
we consulted practitioners from a variety of companies.
Table~\ref{participant_table} presents the characteristics of the participants
that accepted our invitation.
To maintain anonymity, in conformance with the human ethics guidelines,
hereafter we will refer to the participants as P1--P15 (first column). \emph{We
assumed a non-disclosure agreement with the investigated companies to use the
data only in the context of our study and, therefore, we can not disclose them}.

\begin{table}[t]
\centering
\caption{Participant Profile. SX means software development experience in years,
DX means DevOps experience in years, CN means country of work, and CS means
company size (S\textless100; M\textless1000; L\textless5000; XL\textgreater5000).}
\label{participant_table}
\begin{tabular}{p{0.4cm}p{2.6cm}p{0.4cm}p{0.45cm}p{0.5cm}p{1.3cm}p{0.3cm}} \toprule \centering
\textbf{P\#}          & \textbf{Job Title}
       & \textbf{SX} & \textbf{DX} & \textbf{CN}   & \textbf{Domain}    & \multicolumn{1}{l}{\textbf{CS}} \\ \midrule \centering
P1                   & DevOps Developer      & 9            & 2           & IR            & IT                 & S                               \\ \centering

P2                   & DevOps Consult.       & 9            & 3           & BR            & IT                 & M                               \\ \centering

P3                   & DevOps Developer      & 8            & 1           & IR            & IT                 & S                               \\ \centering

P4                   & Computer Tech.        & 10           & 2           & BR            & Health             & S                               \\ \centering

P5                   & Systems Engineer      & 10           & 3           & SP            & Telecom            & XL                              \\ \centering

P6                   & Developer             & 3            & 1           & PO            & IT                 & S                               \\ \centering

P7                   & Support Analyst       & 15           & 2           & BR            & Telecom            & L                               \\ \centering

P8                   & DevOps Engineer       & 20           & 9           & BR            & Marketing              & M                               \\ \centering

P9                   & IT Manager            & 14           & 8           & BR            & IT                 & M                               \\ \centering

P10                  & Network Admin.        & 15           & 3           & BR            & IT                 & S                               \\ \centering

P11                  & DevOps Superv.                & 6            & 4           & BR            & IT                  & M                               \\ \centering

P12                  & Cloud Engineer              & 9            & 3           & US            & IT                  & L                               \\ \centering

P13                  & Technology Mngr.                 & 18            & 6           & BR            & Food                  & M                               \\ \centering

P14                  & IT Manager            & 7            & 2           & BR            & IT                  & S                               \\ \centering

P15                  & Developer        & 3            & 2           & BR            & IT                  & S \\ \bottomrule
\end{tabular}
\end{table}

The interviews were conducted between April 2017 and April 2018 by means of Skype calls
with minimum duration of 20 minutes, maximum of 50 and an average of 31.
Data collection and analysis were iterative so the collected data helped to guide
future interviews. Questions evolved according to
the progress of the research. We started with five open-ended questions: (1) What
motivated the adoption of DevOps? (2) What does DevOps adoption mean in the context of
your company? (3) How was DevOps adopted in your company? (4) What were the
results of adopting DevOps? And (5) what were the main difficulties?

As the analyzes were being carried out, new questions were added to the script.
These new questions were related to the concepts and categories identified in
previous interviews. Examples of new questions include: (1) What is the
relationship between deployment automation and DevOps adoption? (2) Is it
possible to adopt DevOps without automation? (3) How has your company fostered a
collaborative culture?

With respect to \emph{data analysis}, the interviews were
recorded, transcribed, and analyzed. The interviews with participants from
Brazil and Portugal were translated from Portuguese into
English. The first moment of the analysis, called open coding in GT, starts
immediately after the transcription of the first interview.
Open coding lasted until there was no
doubt about the core category of the study. Similar to that described by
Adolph et al.~\cite{adolph2012reconciling}, we started
considering a core category candidate and changed later. The first core category
candidate was \cat{automation}, but we realized that this category did not
explain most of the behaviors or events in data. The sense of
shared responsibilities in solving problems, and the notion of product thinking
are examples of events that could not be naturally explained around \cat{automation}.
We then started to understand that \cc also appeared recurrently in the analysis
and with more potential to explain the remaining events. Thus, we asked explicitly
about the role of \cat{automation} and how the \cc is formed
in a DevOps adoption process.

Considering the script adaptations and the analysis of new data in a constant
comparison process, taking into account the previous analyses and the
respective memos written during all the process, after the tenth
interview, we concluded that \cc was unequivocally the core
category regarding how DevOps was successfully adopted.
At this moment, the open coded ended and the selective coding started.
We started by restricting the coding only
to specific variables that were directly related to the core category and their
relationships. Following three more interviews and respective analysis, we realized that
the new data added less and less content to the emerging theory. That is, the
explanation around how the \cc category is developed showed signs of saturation.
We then conducted two more interviews to conclude that we had reached a
theoretical saturation, that is, we were convinced there were no more enablers
or outcomes related to DevOps adoption, the relationship between all of them
was adequate and the properties of core category were well developed.

At this point, we started the theoretical coding to find a way to integrate
all the concepts, categories, and memos in the form of a cohesive and
homogeneous theory, where we have pointed out the role of the categories as
enablers and outcomes. We will present more details about
the results of our theoretical coding phase in the next section.
To illustrate the coding procedures, we will show a working example from an
interview transcription to a category.
It is important to note that \emph{raw interview transcripts} are full of noise.
We started the coding by removing this noise and identifying the key points.
Key points are summarized points from sections of the interview~\cite{georgieva2008best}.
For example:

\textbf{Raw data:} \textit{``So, here we have adopted this type of strategy that is
the infrastructure as code, consequently we have the versioning of our entire
infrastructure in a common language, in such a way that any person, a
developer, an architect, the operations guy, or even the manager, he can look
at it and describe that the configuration of application x is y. So, it
aggregates too much value for us exactly with more transparency''}

\textbf{Key point:} \textit{``Infrastructure as code contributes to
transparency because it enables the infrastructure versioning in a common
language to all professionals''}

We then assigned codes to the key point. A code is a phrase that summarizes
the key point and one key point can lead to several codes \cite{hoda2017becoming}.

\textbf{Code:} \textit{Infrastructure as code contributes to transparency}

\textbf{Code:} \textit{Infrastructure as code provides a common language}

In this example, the concept that emerged was ``infrastructure as code''. The
expression corresponding to this concept comes directly from raw data, but this
is not a rule. It is common for the concept to be an abstraction, without
emerging from an expression present in raw data.
At this moment, we already identified other concepts that
contribute to transparency. We wrote the following memo:

\textbf{Memo:} \textit{Similar to sharing on a regular basis and shared
pipelines, the concept of infrastructure as code is an important transparency
related one. These transparency related concepts have often been cited as
means to achieve greater collaboration between teams}.

The constant comparison method was repeated on the concepts to produce a third
level of abstraction called categories. Infrastructure as code was grouped
together with five other concepts into the \textbf{sharing and transparency} category.

\section{Categories and Concepts} \label{sec:categories_concepts}

Here we detail our
understanding of the core category
of DevOps adoption (\cc)
and relate it to categories that either work as DevOps {\bf enablers} or
are expected {\bf outcomes} of a DevOps adoption process.
We have highlighted the concepts along with raw data quotes from the interviews.

\subsection{The Core Category: Collaborative Culture}

The \cc is the core category
for DevOps adoption. A \cc essentially aims to remove
the silos between development and operations teams and activities.
As a result, operations tasks---like deployment, infrastructure provisioning
management, and monitoring--- should be considered as regular, day-to-day,
development activities. This leads to the first concept related to
this core category: {\bf operations tasks should be performed by
the development teams in a seamless way.}

\begin{mq}
``\emph{A very important step was to bring the deployment into day-to-day
development, no waiting anymore for a specific day of the week or month. We wanted
to do deployment all the time. Even if in a first moment it were not in
production, a staging environment was enough. [...] Of course, to carry out the deployment
continuously, we had to provide all the necessary infrastructure at the same
pace.}'' (P14, IT Manager, Brazil)
\end{mq}

Without DevOps, a common scenario is an accelerated software development
without concerns about operations. At the end, when the development team has a
minimum viable software product, it is sent to the operations team for
publication. Knowing few things about the nature of the software and how it
was produced, the operations team has to create and configure an environment
and to publish the software. In this scenario, software delivery is typically
delayed and conflicts between teams show up. When a \cc is fomented, teams collaborate to perform the tasks from the first day
of software development. With the constant exercise of provisioning, management,
configuration and deployment practices, software delivery becomes more natural,
reducing delays and, consequently, the conflicts between teams.

\begin{mq}
``\emph{We work using an agile approach, planning 15-day sprints where we focused on
producing software and producing new releases at a high frequency. However, at the time of
delivering the software, complications started to appear. (...) Deliveries often
delayed for weeks, which was not good neither for us nor for stakeholders.}''
(P6, Developer, Portugal)
\end{mq}

As a result of constructing a \cc, the development
team no longer needs to halt its work waiting for the creation
of one application server, or for the execution of some database script, or for
the publication of a new version of the software in a staging environment.
Everyone needs to know the way this is done and, with the collaboration of the
operations team, this can be performed in a regular basis. If any task can be
performed by the development team and there is trust between the teams, this task is
incorporated into the development process in a natural way, manifesting the
second concept related to \cc category: \textbf{software
development empowerment}.

\begin{mq}
``\emph{
It was not feasible to have so many developers generating artifacts and
stopping their work to wait for another completely separate team to publish it. Or
needing a test environment and having to wait for the operations team to
provide it only when possible. These activities have to be available to quickly
serve the development team. With DevOps we supply the need for freedom and
have more power to execute some tasks that are intrinsically linked to their work.}''
(P5, Systems Engineer, Spain)
\end{mq}

A \cc requires  \textbf{product thinking}, in substitution to
\textbf{operations or development thinking}. The development team has to understand that
the software is a product that does not end after ``pushing'' the code to a
project's repository and the operations team has to understand that its
processes do not start when an artifact is received for publication. \textbf{Product thinking}
is the third concept related to our core category.

\begin{mq}
``\emph{We wanted to
hire people who could have a product vision. People who could see the
problem and think of the best solution to it, not only thinking of a
software solution, but also the moment when that application will be
published. We also brought together developers to reinforce that everyone
has to think of the product and not only in their code or in their
infrastructure}'' (P12, Cloud Engineer, United States)
\end{mq}

There should be a \textbf{straightforward communication} between teams. Ticketing
systems are cited as a typical and inappropriate means of communication
between development and operations teams. Face-to-face communication is the best
option, but considering that it is not always feasible, the continuous use of
tools like \emph{Slack} and \emph{Hip Chat} was cited as appropriate options.

\begin{mq}
``\emph{We also use this tool (Hip Chat) as a way to facilitate communication between
development and operations teams. The pace of work there is very accelerated, and thus
it is not feasible to have a bureaucratic communication. (...) This gave us a lot of
freedom to the development activities, in case of any doubt, the operations staff
is within the reach of a message.}'' (P5, Systems Engineer, Spain)
\end{mq}

There is a \emph{shared} responsibility to identify and fix the issues
of a software when transitioning to production. The strategy of avoiding liability should be kept away.
The development team must not say that a given issue is a problem in the infrastructure, then
it is operations team' responsibility. Likewise, the operations team
must not say that a failure was motivated by a problem in the application, then it is
development team's responsibility. A \textbf{blameless} context must exist.
The teams need to focus on solving problems, not on laying the blame on others
and running away from the responsibility. The sense of \textbf{shared
responsibilities} involves not only solving problems, but also any other
responsibility inherent in the software product must be shared.
\textbf{Blameless} and \textbf{shared responsibilities} are the remaining
concepts of the core category.

\begin{mq}
``\emph{We realized that some people were afraid of making mistakes. Our
culture was not strong enough to make everyone feel comfortable to innovate and
experiment without fear of making mistakes. We made a great effort to spread
this idea that no-one is to be blamed for any problem that may occur. We take
every possible measure to avoid failures, but they will happen, and only without
blaming others we will be able to solve a problem quickly.}'' (P8, DevOps Engineer, Brazil)
\end{mq}

At first glance, considering the creation and strengthening of the \cc as the most important step towards DevOps adoption seems somewhat obvious, but
the respondents cited some mistakes that they consider recurrent in not
prioritizing this aspect in a DevOps adoption:

\begin{mq}``\emph{In a DevOps adoption, there is a very strong cultural issue that the teams
sometimes are not adapted to. Regarding that, one thing that bothers me a lot
and that I see very often is people hitching DevOps exclusively by tooling or
automation.}'' (P9, IT Manager, Brazil)
\end{mq}

Besides the core category (\cc), we have identified
three other sets of categories: the enablers
of DevOps adoption, the consequences of adopting
DevOps, and the categories that are both enablers and consequences.

\subsection{Enabler Categories}

Below, we detail the categories that support the adoption of
DevOps practices, including \cat{automation}, \cat{sharing and transparency}.

\subsubsection{Automation} \label{ssec:automation}

This category presents the higher number of related concepts. This
occurs because manual proceedings are considered strong candidates to
propitiate the formation of a silo, hindering the construction
of a \cc. If a task is manual, a single person or
team will be responsible to execute it. Although \cat{transparency} and \cat{sharing} can
be used to ensure collaboration even in manual tasks, with automation the
points where silos may arise are minimized.

\begin{mq}
``\emph{When a developer needed to build a new application, the previous workflow demanded him
to create a ticket to the operations teams, which should then manually evaluate and solve
the requested issue. This task could take a lot of time and there was no
visibility between teams about what was going on (\ldots). Today, those silos do not exist
anymore within the company, in particular because it is not necessary to execute all these tasks manually.
Everything has been automated.}'' (P12, Cloud Engineer, United States)
\end{mq}

In addition to contributing to \cat{transparency}, \cat{automation} is also considered
important to ensure \emph{reproducibility} of tasks, reducing rework and risk of
human failure. Consequently, \cat{automation} increases the confidence
between teams, which is an important aspect of the \cc.

\begin{mq}
``\emph{Before we adopted DevOps, there was a lot of manual work. For example, if you
needed to create a database schema, it was a manual process; if you needed to create a
database server, it was a manual process; if you needed to create additional EC2 \footnote{Amazon Elastic
Compute Cloud} instances, such a process was also manual.
This manual work was time consuming and often caused errors and
rework.}'' (P1, DevOps Developer, Ireland)
\end{mq}

The eight concepts of the \cat{automation} category will be detailed next.
In all interviews we extracted explanations about \textbf{deployment
automation} (1), as part of DevOps adoption. Software delivery is the clearest
manifestation of value delivery in software development. In case of problems
in deployment, the expectation of delivering value to business can quickly
generate conflicts and manifest the existence of silos.
In this sense, \cat{automation} typically increases agility and reliability. Some other
concepts of automation go exactly around deployment automation.

It is important to note that frequent and successfully
deployments are not sufficient to generate value to business. Surely, the quality of
the software is more relevant. Therefore, quality checks need to be automated as well, so they can be part of the
deployment pipeline, as is the case of \textbf{test automation} (2). In addition, to
automate application deployment, the environment where the
application will run needs to be available. So, \textbf{infrastructure
provisioning automation} (3) must be also considered in the process. Besides being available,
the environment needs to be properly configured, including the amount of memory and CPU,
availability of the correct libraries versions, and database structure. If the configuration of some of these concerns
has not been automated, the deployment activity can go wrong. Therefore,
the automation of \textbf{infrastructure management} (4) is another
concept of the \cat{automation} category.

Modern software is built around services. Microservices  was commonly cited
as one aspect of DevOps adoption. To Fowler and Lewis
\cite{martinfowler2014microservices}, in the
microservice architectural style, services need to be independently deployable
by fully automated deployment machinery. We call this part of microservices
characteristics of \textbf{autonomous services} (5). \textbf{Containerization}
(6) is also mentioned as a way to automate the provisioning of containers---the
environment where these autonomous services will execute.
\textbf{Monitoring automation} (7) and \textbf{recovery automation} (8) are the
remaining concepts. The first refers to the ability to monitor the
applications and infrastructure without human intervention. One classic example
is the widespread use of tools for sending messages reporting
alarms---through SMS, Slack/Hip Chat, or even
cellphone calls-- in case of incidents. And the second is related to the ability
to either replace a component that is not working or
roll back a failed deployment without human intervention.

\subsubsection{Transparency and Sharing} It represents the grouping of concepts
whose essence is to help disseminate information and ideas among all. Training,
tech talks, committees lectures, and round tables
are examples of these events. Creating
channels by using communication tools is another recurrent topic
related to \cat{sharing} along the processes of DevOps adoption.
According to the content of what is shared, we have identified three main concepts:
(1) {\bf knowledge sharing}: the professionals interviewed mention a wide range of
skills they need to acquire during the adoption of DevOps, citing
structured sharing events to smooth the learning curve of both technical and
cultural knowledge; (2) {\bf activities sharing}: where the focus is on sharing how simple tasks can or
should be performed (e.g., sharing how a bug has been solved). Communication tools,
committees, and round tables are the common forum for sharing this type of content;
and (3) {\bf process sharing}: here, the focus is on sharing whole working processes
(e.g., the working process used to provide a new application server). The
content is more comprehensive than in sharing activities. Tech talks and
lectures are the common forum for sharing processes.

Sharing concepts contribute with the \cc. For example,
all team members gain best insight about the entire software production
process, with a solid understanding of shared responsibilities. A shared vocabulary also
emerged from \cat{sharing} and this facilitates communication.

The use of \textbf{infrastructure as code} was
recurrently cited as a means for guaranteeing that everyone knows how the execution environment of
an application is provided and managed. Bellow, is an interview
transcript which sums up this concept.

\begin{mq}
``\emph{So, here we have adopted this type of strategy that is the infrastructure as code,
consequently we have the versioning of our entire
infrastructure in a common language, in such a way that any person, a
developer, an architect, the operations guy, or even the manager, he can look
at it and describe that the configuration of application x is y. So, it
aggregates too much value for us exactly with more transparency.}'' (P12, Cloud Engineer, United States)
\end{mq}

Regarding \cat{transparency and sharing}, we have also found the concept of
\textbf{sharing on a regular basis}, which suggests that sharing should be
embedded in the process of software development, in order to contribute
effectively to transparency (e.g., daily meetings with Dev and Ops staff
together was one practice cited to achieve this).
As we will detail in the \emph{continuous integration} concept of
the \cat{agility} category, a common way to integrate all tasks is a pipeline.
Here, we have the
concept of \textbf{shared pipelines}, which indicates that the code of pipelines
must be accessible to everyone, in order to foment transparency.

\begin{mq}
``\emph{The code of how the infrastructure is
made is open to developers and the sysadmins need to know some aspects of how
the application code is built. The code of our pipelines is accessible to
everyone in the company to know how activities are automated}'' (P13, Technology
Manager, Brazil)
\end{mq}

\subsection{Categories related to the DevOps adoption outcomes}

In this section we detail the categories that correspond to
the expected consequences with the adoption of
DevOps practices, including \cat{agility} and \cat{resilience};
as discussed as follows.

\subsubsection{Agility}

Agility is frequently discussed as a major outcome of DevOps adoption. With more
collaboration between teams, \textbf{continuous integration} with the execution of
multidisciplinary pipelines is possible, and it is an agile related concept
frequently explored. These pipelines might contain
infrastructure provisioning, automated regression testing, code analysis,
automated deployment and any other task considered important
to continuously execute.

These pipelilnes encourage two other agile concepts: \textbf{continuous
infrastructure provisioning} and \textbf{continuous deployment}. The latter is
one of the most recurrent concepts identified in the interview analysis. Before
DevOps, deployment had been seen as a major event with high risk of downtime and
failure involved. After DevOps, the sensation of risk in deployment decreases
and this activity became more natural and frequent. Some practitioners claim
to perform dozens of deployments daily.

\subsubsection{Resilience}

Also related to an expected outcome of adopting DevOps, \cat{resilience} refers
to the ability of applications to adapt quickly to adverse situations.
The first related concept is \textbf{auto scaling}---i.e.,
allocating more or fewer resources to applications that increase or
decrease on demand. Another concept related to
the \cat{resilience} category is \textbf{recovery automation}, that is
the capability that applications and infrastructure have to recovery itself in case of
failures. There are two typical cases of recovery automation: (1) in cases
of instability in the execution environment of an application (a
container, for example) an automatic restart of that environment will occur; and (2) in
cases of new version deployment, if the new version does not work properly, the
previous one must be restored. This auto restore of a previous version
decrease the chances of downtimes due to errors in specific versions, which
is the concept of \textbf{zero down-time}, the last one of the \cat{resilience} category.

\subsection{Categories that are both Enablers and Outcomes}

Finally, we will detail bellow the categories that are both enablers
and outcomes, including \cat{continuous measurement}
and \cat{quality assurance}; as discussed as follows.

\subsubsection{Continuous Measurement}

As an enabler, regularly performing the measurement and sharing activities
contributes to avoiding existing silos and reinforces the \cc, because it is
considered a typical responsibility of the operations team.

\begin{mq}
``\emph{Before, we had only sporadic looks to
zabbix\footnote{\url{https://www.zabbix.com/}} to check if everything was OK.
At most someone would stop to look memory and CPU consumption. To maintain
the quality of services, we expanded this view of metrics collection so that it
became part of the software product. We then started to collect metrics continuously
and with shared responsibilities. For example, if an overflow occurred in the
number of database connections, everyone received an alert and had
the responsibility to find solutions to that problem.}'' (P3, DevOps Developer, Ireland)
\end{mq}

As an outcome, the continuously collection of metrics from applications and
infrastructure was appointed as a necessary behavior of the teams after the adoption
of DevOps. It occurs because the resultant agility increases the risk of
something going wrong. The teams should be able to react quickly in case of
problems, and the continuous measurement allows it to be proactive and resilient.

\begin{mq}
``\emph{With DevOps we can do deployment all the time and, consequently, there was
a need for greater control of what was happening. So, we used
grafana\footnote{\url{https://grafana.com/}} and
prometheus\footnote{\url{https://prometheus.io/}} to follow everything that is
happening in the infrastructure and in the applications. We have a complete
dashboard in real time, we extract reports and, when something goes wrong, we
are the first to know.}'' (P10, Network Administrator, Brazil)
\end{mq}

Continuous monitoring involves \textbf{application log monitoring} (1), a
concept that corresponds to the use of the log produced by
applications and infrastructure as data source. The concept of
\textbf{continuous infrastructure monitoring} (2) indicates that the monitoring
is not performed by a specific person or team in a specific moment. The
responsibility to monitor the infrastructure is shared and it is executed on
a daily basis. \textbf{Continuous application measurement} (3), in turn, refers to
the instrumentation to provide metrics that are used to evaluate aspects and
often direct evolution or business decisions. All this monitoring/measurement
can occur in an automated way, the \textbf{monitoring automation} has already been
detailed in subsection \ref{ssec:automation}.

\subsubsection{Quality Assurance}

In the same way as continuous measurement, quality assurance is a category that
can work both as enabler and as outcome. As enabler because increasing quality
leads to more confidence between the teams, which in the end generates a virtuous
cycle of collaboration. As outcome, the principle is that it is not
feasible to create a scenario of continuous delivery of software with no control
regarding the quality of the products and its production processes.

Respondents pointed to the need for a sophisticated control of which code should
be part of deliverables that are continuously delivered. Git Flow was
recurrently cited as a suitable \textbf{code branching} (1) model, the first
concept of quality assurance.
In a previous section, we explored the automation face of
microservices and testing. These elements have also a quality assurance face.
Another characteristic of microservices is the need for small services focusing
in doing only one thing. These small services are easier to scale and
structure, which manifest a quality assurance concept: \textbf{cohesive
services} (2). Regarding testing, another face is \textbf{continuous
testing} (3). To ensure quality in software products, we found that
tests (as well as other quality checks) should occur continuously. Continuous testing
is considered challenging without automation, and this reinforces the need for automated
tests.

Another two concepts cited as part of quality assurance in DevOps adoption are
the use of \textbf{source code static analysis} (4) to compute quality metrics in
source code and the \textbf{parity between environments} (development, staging
and production) to reinforce transparency and collaboration during software
development.

\section{A Theory on DevOps Adoption} \label{sec:results}

The results of a grounded theory study, as the name of the method itself
suggests, are grounded on the collected data, so the hypotheses emerge from
data. A grounded theory should describe the key relationships between the
categories that compose it, i.e., a set of inter-related hypotheses~\cite{hoda2017becoming}.
We present the categories of our grounded theory
about DevOps adoption as a network of the two categories of enablers (\cat{automation},
\cat{sharing and transparency}) that are commonly used to develop the core category
\cc, as discussed in the previous section. Based on our understanding,
implementing the enablers to develop the \cc typically leads
to concepts related to two categories of expected outcomes:
\cat{agility} and \cat{resilience}. Moreover, there are two categories that can be considered
both as enablers and as outcomes: \cat{continuous measurement} and \cat{quality assurance}.
In this section, we describe the relationships between those categories, building a theory
of DevOps adoption.

\subsection{A General Path for DevOps Adoption}

In Section~\ref{sec:introduction} we presented the general question of this
research: is there any recommended path to adopt DevOps? Here, we elaborated a response,
based on the analyses conducted as detailed in Section~\ref{sec:research_method}. The main
point that should be formulated is the construction of a \cat{collaborative
culture} between the software development and operations teams and
related activities. According to our findings, the other categories,
many of which are also present in other studies that have investigated DevOps,
only make sense if the practices and
concepts related to them either contribute to the level of a \cc or lead to the expected consequences
of a \cc. This understanding induces several hypothesis, as shown below.

\begin{mh}
\textbf{Hypothesis 1:} \textit{Certain categories related to DevOps adoption
only make sense if used to increase the} \cc \emph{level. We
call this set of categories of \textbf{enablers}}.
\end{mh}

Based on this first hypothesis, the maturity of DevOps adoption does not
advance in situations where only one team is responsible to understand, adapt, or
evolve automation---even when such automation supports different activities like
deployment, infrastructure provisioning and monitoring. The same holds for the
other \emph{enabling} categories. That is, in situations that
\cat{transparency and sharing} do not contribute to
the \cc, they do not contribute to DevOps adoption as a whole. Some examples
that support our first hypothesis include:

\begin{mq}
``\emph{DevOps involves tooling, but DevOps is not tooling. That is, people often
focus on using tools that are called `DevOps tools', believing that this is
what DevOps is. I always insist that DevOps is not tooling, DevOps involves the
proper user of tools to improve software development procedures.}'' (P2, DevOps
Consultant, Brazil)
\end{mq}

\begin{mh}
\textbf{Hypothesis 2:} \textit{Some other categories are not related to DevOps
adoption for contributing to increase the} \cc \emph{level, but for emerging
as an expected or necessary consequence of the adoption. These categories
represent the set of \textbf{outcomes}}.
\end{mh}

In a first moment, the simple fact that a team is more
\cat{agile} in delivering software, or more \cat{resilient} in failure recovery, does not
contribute directly to bringing operations teams closer to development teams.
Nevertheless, a signal of a mature DevOps adoption is an increased capacity for continuously
delivering software (and thus being more \cat{agile})
and for building \cat{resilient} infrastructures.

\begin{mh}
\textbf{Hypothesis 3:} \textit{The categories \cat{Continuous Measurement} and \cat{Quality Assurance}
are both related to DevOps enabling capacity and to DevOps outcomes}.
\end{mh}

Measurement is cited as a typical responsibility of the operations team.
At the same time that sharing this responsibility reduces silos,
it is also cited that measurement is a necessary consequence of DevOps adoption. Particularly because
the continuous delivery of software requires more control,
which is supplied by concepts related to the \cat{continuous measurement} category.
The same premise is valid to the \cat{quality assurance} category. At first glance,
\cat{quality assurance} appears as one response to the context of agility in operations
as a result of DevOps adoption. But, the efforts in quality assurance of software products
increase the confidence between the development and operations teams, increasing the level
of \cc.

Altogether, DevOps enablers are the means commonly used to increase the level of
the \cc in a DevOps adoption process.
We have identified five categories of DevOps enablers:
\cat{ Automation}, \cat{Continuous Measurement}, \cat{Quality Assurance},
\cat{Sharing}, and {\cat{Transparency}. Another finding of our
study leads to our fourth hypothesis.

\begin{mh}
\textbf{Hypothesis 4:} \textit{There is no precedence between enablers in a DevOps adoption process}.
\end{mh}

We have realized that the adoption process might not have
to priorize any enabler, and a company that aims to implement
DevOps should start with  the enablers that seem more appropriate (in terms
of its specificities). Accordingly, we did not find any evidence that an enabler
is more efficient than another for creating a \cc. \cat{Automation} is the category
that appears more frequently in our study, though several participants make
clear that associating DevOps with automation is a misconception.

\begin{mq}
``\emph{I think that the expansion of collaboration between teams involved other
things. It was not just automation. There must be an alignment with the
business needs. (...) I think that DevOps enabled a broader understanding
of software production and we realized the very fact that it is not about
automating everything. (...) So, I see with caution a supposed vision that
automating things can be the way to implement DevOps.}'' (P7, Support Analyst, Brazil)
\end{mq}

DevOps outcomes are the categories that does not primarily produce the
expected effect of an {\bf enabler}, typically concepts that are expected as
consequences of an adoption of DevOps. We have identified four categories that
can work as DevOps outcomes: \cat{agility}, \cat{continuous measurement},
\cat{quality assurance}, and \cat{software resilience}. Note that,
as mentioned before, \cat{continuous measurement} and \cat{quality assurance}
are both enablers and outcomes.

That is, a well succeeded DevOps adoption typically increases the potential of
\cat{agility} of teams and enables \cat{continuous measurement}, \cat{quality assurance} and
\cat{resilience} of applications.
However, in some situations, this potential is not completely used due to business
decisions. For example, one respondent cited that, at a first moment, the
company did not allow the continuous deployment (more potential of agility)
of applications in production:

\begin{mq}
``\emph{We had conditions and security to continuously publish in production,
however, in the beginning, the managers were afraid and decided that the
publication would happen weekly.}'' (P9, IT Manager, Brazil)
\end{mq}

\subsection{A Model for DevOps Adoption and Its Application}\label{sec:case_study}

Based on H1-H4 hypothesis, we present a three step model that
explains how to adopt DevOps according to our understanding. The
model considers the following steps:

\begin{enumerate}
\item In the first step, a company should
disseminate that the goal with a DevOps adoption is to
establish a \cc between
development and operations teams.

\item In the second step, a company should select and develop
the most suitable enablers according to its context. The enablers
are means commonly used to develop the \cc
and its concepts.

\item In the third step, a company should check the outcomes of the
DevOps adoption in order to verify the alignment with
industrial practices and to explore them according to the company's need.
\end{enumerate}

Our proposed model has been applied to guide the DevOps adoption at the Brazilian Federal Court of
Accounts (TCU) where one of the authors of this study works as a software
developer. The TCU is responsible for the accounting, financial, budget, performance, and property
oversight of federal institutions and entities of the country. Currently, there are 2500
professionals working at the TCU, of which approximately 300 work directly on either
software development or operations. The source code repository at the TCU hosts more then 200 software projects, totaling
over 4 million lines of code.
Before the application of our model, the TCU had produced some w.r.t deployment
automation results and the focus was being directed to the tooling issue. Considering this
incomplete perspective of DevOps, the conflicts between development and operations
teams continued. That is, the mere advance in implanting ``DevOps tools'' simply
changed the points of conflict, but they persisted.

After the presentation of our  model in a series of lectures, development and
operations teams changed their focus to build a \cc. This
change was only possible due to the engagement and sponsorship of the IT
managers. Looking to the concepts within the \cc category, the first practical
action at the TCU was to facilitate communication between teams. The use of tickets
was then abolished. The problems had to be solved in a collaborative way, preferably
face to face.
Looking to enablers, the TCU is applying \cat{sharing and transparency} concepts.
The role of internal tech talks and committees to disseminate that collaboration
culture and related concepts is being reinforced.
When a new infrastructure had to be provided and configured, the current guideline is
that there must be a kind of \emph{pair programming} between developers and infrastructure
members. All application related tasks must be executed in a collaborative
way. Naturally, the professionals noticed that automation would facilitate the
operationalization of that collaboration. For this reason, the infrastructure provisioning
and management was automated.

The TCU also uses continuous measurement and quality assurance concepts as
enablers of its DevOps adoption. The applications started to be continuously
tested and measured. The tests were automated and included in the pipelines.
Verification of test coverage and quality code also became part of the pipeline.
This increased the confidence between teams. The TCU started
to explore the potential of DevOps tools, like recovery automation, zero
down-time, and auto scaling. The deployment has also been automated.
It is important to note that, before DevOps, deployment activities were historically a controversial point at the TCU.
Several conflicts occurred over time. Rigid procedures were created to try to
avoid problems. These ``rigid procedures'' often led to periods of months
without any software delivery. The more collaborative scenario, with a strong appeal in automation and quality,
created by following an appropriate path in adopting DevOps, enabled the deployment activities to become
a lightweight task at the TCU. Continuous deployment became a reality and, currently, several deployments
occur as regular activities of the development teams at the TCU.

Since the TCU is a government institution, some advances in DevOps adoption still comes up
against regulatory issues. For example, there are internal regulations that
establish that only the operations sector is responsible for issues related to
application infrastructure, contrasting with shared responsibilities that are
part of the \cc. Nevertheless, our model enabled the TCU to adopt DevOps in a more
sustainable way. Knowing the role
of each DevOps element in the adoption was fundamental for the TCU to avoid points
of failure and to build a collaborative environment that supports the
exploration of DevOps benefits.

\section{Threats to Validity}

Regarding construct validity,  we are actually relying on the subjective
practitioners' perception when we stated that we performed our study considering successful cases
of DevOps adoptions. However, currently, there is no objective way to measure whether or not a
DevOps adoption was successful.
Although Grounded Theory offers rigorous procedures for data analysis, our
qualitative research may contain some degree of research bias. Certainly, other
researchers might form a different interpretation and theory after analyzing
the same data, but we believe that the main perceptions would be preserved.
This is a typical threat related to GT studies, which do not claim to generate
definitive findings. The resulting theory, for instance, might
be different in other contexts \cite{hoda2012developing}.

For this reason, we do not claim
that our theory is absolute or final. We welcome extensions to the theory based
on unseen aspects or finer details of the present categories or potential discovery
of new dimensions from future studies.
Future work can also focus on investigating contexts
where DevOps adoption did not succeed, aiming to validate if our model could be
relevant in this scenario too. Finally, regarding external validity, although we
considered in our study the point of view of practitioners with different
backgrounds, working in companies from different domains, and distributed across
five countries, we do not claim that our results are valid for
other scenarios---although we almost achieved saturation
after the 12$^{th}$ interview. Accordingly, our degree of heterogeneity complement
previous studies that mostly focus in a single company (as we will discuss next).

\section{Related Work} \label{sec:related_work}

A literature review by Erich et al.~\cite{cooperation_dev_ops_esem_2014} presents 8
main concepts related to DevOps: culture, automation, measurement, sharing,
services, quality assurance, structures and standards. The authors pointed out
that the first four concepts are
related to the CAMS framework, proposed by Willis~\cite{what_devops_means_2010}.
The paper concludes that there is a great opportunity for empirical researchers
to study organizations experimenting with DevOps.
Other studies (e.g.,~\cite{devops_a_definition_xp_15,dimensions_of_devops_xp_15,extending_dimensions_icsea_16,characterizing_devops_sbes_2016,qualitative_devops_journalsw_17})
mixed literature reviews with empirical data to investigate DevOps.
Although our research and recent literate are interested in understanding DevOps,
there are subtle differences in both (1) the methodological aspects and (2) the focus
of each work.

First of all, none of the aforementioned works focused on explaining the process of DevOps adoption,
in particular, using data collected in the industry. This is unfortunate, since the
practitioners' perception present an unique point of view that researchers
alone could hardly grasp. Moreover, although the literature has a number of
useful elements, there is a need to complement such elements with a perspective on how DevOps has
been adopted, containing guidance about how to connect all these isolated parts
and then enabling new candidates to adopt DevOps in a more consistent way.
For instance, the work of Erich et al.~\cite{qualitative_devops_journalsw_17}
focus on investigating the ways in which organizations implement DevOps.
However, this work relies only in literature review and does not formulate
new hypothesis about DevOps adoption. Second,
in terms of results, our main distinct contribution is to improve the guidance
to new practitioners in DevOps adoption.
Next, we present the overlappings of our
results with the existing literature, presenting also the main differences that
make the contributions of our work clearer.

The work of J. Smeds et al.~\cite{devops_a_definition_xp_15} uses a literature
review to produce one explanation about DevOps through a set of enablers and capabilities. Additionally, their results
present a set of impediments of DevOps adoption based on an interview with 13
subjects of a same company, and whose DevOps adoption process was at
an initial stage. The main similarities with our study are: (1) grouping
elements as DevOps enablers; and (2) the presence of several similar concepts:
(a) testing, deploying, monitoring, recovering and infrastructure automation;
(b) continuous integration, testing and deployment; (c) service failure recovery
without delay; and (d) constant, effortless communication. The main
differences are: (1) their work does not group concepts into categories,
for example: most of their enablers were grouped together by us within the \cat{automation} category; (2) presents cultural enablers as
common contributor to DevOps, not as the most important concern; and (3) the empirical
part of the study focus on building a list of possible impediments to DevOps
adoption, not on providing guidance to new adopters.

Lwakatare et al.~\cite{extending_dimensions_icsea_16} proposed a conceptual
framework to explain ``DevOps as a phenomenon''. The framework is organized around
five dimensions (collaboration, automation, culture, monitoring and measurement)
and these dimensions are presented with related practices. The main similarity
with our study is that all dimensions are also presented here. The
main differences are: (1) collaboration and culture are presented by us
as a single abstraction; (2) Concepts related to monitoring and measurement are
grouped by us in a single category: \cat{continuous measurement}; and (3) it does
not indicate a major dimension (aka, the core category).

Fran\c{c}a et al.~\cite{characterizing_devops_sbes_2016} present a DevOps
explanation produced by means of a multivocal literature review. The data was collected
from multiple sources, including gray literature, and analyzed by using procedures
from GT. The results contain a set of DevOps principles, where
there is most of the overlapping with our study. In addition, the paper
presents a definition to DevOps, issues motivating its adoption, required skills,
potential benefits and challenges of adopting DevOps. The main similarities
are: (1) Automation, sharing, measurement and quality assurance are presented as
DevOps categories; and (2) Their social aspects category is similar to our
\cc category. The main differences are: (1) it presents DevOps as a
set of principles, different from enablers and outcomes in our study; and (2) the Leanness
category is not present in our study and the \cat{resilience} category is not present
in theirs; and (3) it does not indicate a core category.

The study conducted by Erich et al.~\cite{qualitative_devops_journalsw_17},
similarly to the others cited above, combined literature review with some
interviews with practitioners. In the literature review part, the papers were
labeled and the similar labels are grouped. The 7 top labels are then presented
as elements of DevOps usage in literature: culture of collaboration, automation,
measurement, sharing, services, quality assurance and governance. After the literature
review, six interviews were conducted in order to obtain evidence of DevOps
adoption in practice. The interviews were analyzed individually and a comparison
between them was made, focusing on problems that organizations try to solve by
implementing DevOps, problems encountered when implementing DevOps and practices
that are considered part of DevOps. The main similarity with our study
is that 5 of their 7 groups are also present in our study (culture of collaboration,
automation, measurement, sharing and quality assurance). The main
differences are: (1) it does not consolidate the practitioners' perspective, but
only compare it with literature review results; and (3) it does not indicate a major group.

\section{Final Remarks} \label{sec:conclusion}

In this paper, grounded in data collected from successfully DevOps adoption
experiences, we present a theory on DevOps adoption, a model of how to adopt
DevOps according to this theory, and a case of applying it in practice.

We found out that the DevOps adoption involves a very specific relationship between
seven categories: \cat{agility}, \cat{automation}, \cc, \cat{continuous
measurement}, \cat{quality assurance}, \cat{resilience}, \cat{sharing and transparency}.
The core category of DevOps adoption is the \cc. Some of the
identified categories (i.e., automation and sharing and transparency) propitiate
the foundation of a \cc. Other categories
(i.e., agility and resilience) are expected consequences of this formation.
Finally, two other categories (i.e., continuous measurement and quality
assurance) work as both foundations and consequences. We call the foundations
categories ``DevOps enablers'', and the consequences categories ``DevOps outcomes''.
Crucially, this model simplifies the understanding of the
complex set of elements that are part of DevOps adoption, enabling it to be
more direct and to offer a lower risk of focusing on wrong things.

We experimented with
this model in real settings, improving the benefits of adopting DevOps
within a government institution that faced many problems with the separation between the
development and operations teams.

\bibliographystyle{ACM-Reference-Format}
\bibliography{references}

\end{document}